\documentclass[12pt,twocolumn]{article}

\usepackage{amsmath,amsfonts,amssymb}
\usepackage{graphicx}
\usepackage{caption}
\graphicspath{{./figures/}}
\usepackage[margin=0.7in]{geometry}
\usepackage[utf8]{inputenc}
\usepackage{float}
\usepackage[
    backend=biber,
    citestyle=science,
    style=chem-rsc,
    maxbibnames=1,
]{biblatex}
\AtBeginBibliography{\footnotesize}
\usepackage{hyperref}
\hypersetup{
    colorlinks = true,
    allcolors=blue,
}
\usepackage{abstract}

\usepackage{titlesec}
\titlespacing\section{0pt}{12pt plus 4pt minus 2pt}{8pt plus 4pt minus 2pt}
\titleformat{\section}[block]{\scshape\filcenter}{}{1em}{}

\usepackage{upgreek}
\usepackage{mathptmx}
\newcommand\tab[1][0.4cm]{\hspace*{#1}}
\newcommand\taba[1][0.14cm]{\hspace*{#1}}
\usepackage{tabularx}
\usepackage{changepage}

\begin{document} 
\twocolumn[{
\begin{flushleft}
\Large\textbf{Bulk dissipation in the quantum anomalous Hall effect}
\end{flushleft}

\begin{flushleft}
{Linsey K. Rodenbach$^{1,2}$, Ilan T. Rosen$^{2,3}$, Eli J. Fox$^{1,2}$, Peng Zhang$^{4}$, Lei Pan$^{4}$, Kang L. Wang$^{4}$, Marc~A.~Kastner$^{1,2,5}$, David Goldhaber-Gordon$^{1,2,6}$}
\end{flushleft}

\begin{flushleft}
\footnotesize{$^1$\textit{Department of Physics, Stanford University, 382 Via Pueblo Mall, Stanford, CA 94305, USA}}\\
\footnotesize{$^2$\textit{Stanford Institute for Materials and Energy Sciences, SLAC National Accelerator Laboratory, 2575 Sand Hill Road, Menlo \taba Park, California 94025, USA}}\\
\footnotesize{$^3$\textit{Department of Applied Physics, Stanford University, 348 Via Pueblo Mall, Stanford, CA 94305, USA}}\\
\footnotesize{$^4$\textit{Department of Electrical Engineering, University of California, Los Angeles, California 90095, USA}}\\
\footnotesize{$^5$\textit{Department of Physics, Massachusetts Institute of Technology, 77 Massachusetts Avenue, Cambridge, MA 02139, USA}}\\
\footnotesize{$^6$To whom correspondence should be addressed; E-mail: \texttt{goldhaber-gordon@stanford.edu}}
\end{flushleft}

\begin{abstract}
Even at the lowest accessible temperatures, measurements of the quantum anomalous Hall (QAH) effect have indicated the presence of parasitic dissipative conduction channels. There is no consensus whether parasitic conduction is related to processes in the bulk or along the edges. Here, we approach this problem by comparing transport measurements of Hall bar and Corbino geometry devices fabricated from Cr-doped (BiSb)$_2$Te$_3$. We identify bulk conduction as the dominant source of dissipation at all values of temperature and in-plane electric field. Furthermore, we observe identical breakdown phenomenology in both geometries, indicating that breakdown of the QAH phase is a bulk process. The methodology developed in this study could be used to identify dissipative conduction mechanisms in new QAH materials, ultimately guiding material development towards realization of the QAH effect at higher temperatures.\\
\end{abstract}
}]

\section{Introduction}
In three dimensional topological insulators, strong spin-orbit interactions give rise to 2D Dirac-like surface states with spin-momentum locking, surrounding a fully gapped 3D bulk. An exchange gap opens in this Dirac spectrum when time-reversal symmetry is broken through introduction of ferromagnetism by magnetic doping~\cite{Chen2010,Yu2010,Checkelsky2012}. The result is a Chern insulating state where a single chiral edge mode connects the gapped 2D Dirac bands. By tuning the Fermi level ($E_F$) into the exchange gap, dissipationless 1D conduction at zero external magnetic field is observed. This is known as the quantum anomalous Hall (QAH) effect. Similar to the quantum Hall (QH) effect, transport measurements of the QAH effect are predicted to yield zero longitudinal resistivity $\rho_{xx} = 0$ and a quantized Hall resistivity $ \rho_{yx} = h/e^2$, where $h$ is Planck's constant and $e$ is the electron charge.

The QAH system is a vehicle for tailored control of the spin and momentum of electrons. As parts of new heterostructures, QAH materials could have applications in dissipationless spintronics~\cite{Jiang2019,Wu2014} and quantum computing~\cite{Chen2018,Zeng2018}. Furthermore, QAH systems could replace the quantum Hall (QH) systems used today as resistance standards. Needing no large magnetic field, QAH-based resistance metrology would be more portable and economical. QAH metrological standards would further allow for simultaneous measurement of a resistance standard alongside a Josephson voltage standard in a single cryostat, something that is not currently possible due to the magnetic fields required for measuring QH devices~\cite{Rigosi2019}.\\
\tab Three dimensional thin-film magnetic topological insulators (3D MTIs) have indeed demonstrated the QAH effect~\cite{Bestwick2015,Chang2013a,Checkelsky2014,Kou2014,Kandala2015,Mogi2015,Chang2015,Grauer2015,Feng2015}. Cr- and V-doped (BiSb)$_2$Te$_3$ (Cr-BST and V-BST) samples have shown Hall resistances equal to $h/e^2$ to part-per-million precision~\cite{Fox2018,Gotz2018}. But these samples are not perfect. In all instances this quantization was accompanied by a finite longitudinal resistivity $\rho_{xx} \neq 0$, even at the lowest temperatures and excitation currents, with rapid increases at higher temperatures and currents. This dissipation hinders technological applications of QAH materials as it demands operation at temperatures and current densities far lower than can be used with typical QH systems.

Despite the high ferromagnetic Curie temperature (approximately 20 K for Cr- and V- doped BST samples \cite{Checkelsky2014,Chang2013a,Li2016,Checkelsky2012,Kou2014,Kandala2015,Mogi2015,Grauer2015,Lee2015,Chang2015}), quantization of the Hall resistivity $\rho_{yx}$ to even 5\% accuracy is typically limited to temperatures below 200 mK, with the exception of films with elaborate five-layer modulation of Cr concentration; these have exhibited $\rho_{yx} = 0.97 h/e^2$ at 2 K \cite{Mogi2015}. In all these cases, the amount of dissipation measured at these temperatures is far greater than one would expect given the large exchange gap, which has been shown to be on the order of tens of meV~\cite{Lee2015,Chen2010,Wray2011} These discrepancies may relate to the presence of residual conduction channels which add dissipation to the chiral edge conduction \cite{Chang2015}. Understanding what dissipative conduction channels are present could enable targeted material improvements to increase the temperature at which QAH can be observed.

One proposed source of dissipation in thin-film MTIs is backscattering in nonchiral quasihelical edge modes~\cite{Kou2014,Wang2013,Chang2015a}. In Cr-BST, these quasihelical edge modes are predicted to exist for all samples with a thickness greater than three quintuple layers (QLs)~\cite{Wang2013}. Yet the QAH phase in Cr-BST requires a minimum thickness of four QLs due to competition between the magnetic exchange and the hybridization between top and bottom surface states \cite{Feng2016}. Were quasihelical edge-modes a significant source of dissipation, Cr-BST-based QAH insulators therefore could not host truly dissipationless transport at any thickness, regardless of material improvements. On the other hand, dissipative conduction arising from disorder-based bulk processes, like variable range hopping~\cite{Chang2013a,Kawamura2017} or thermal activation of bulk and surface carriers~\cite{Bestwick2015}, could plausibly be addressed by reducing disorder. Standard transport measurements of Cr-BST have yet to distinguish whether bulk or edge conduction dominates dissipation~\cite{Fox2018,Cuizu2015}.

Most transport studies of QAH materials have employed Hall bar geometries, in which components of the resistivity tensor $\rho$ are extracted straightforwardly. Components of the conductivity tensor $\sigma$ are then computed using the relations $\sigma_{xx} = \rho_{xx} /\left(\rho_{yx}^2 + \rho_{xx}^2 \right)$ and $\sigma_{xy} = \rho_{yx} /\left(\rho_{yx}^2 + \rho_{xx}^2 \right)$. In the Hall bar geometry, both bulk and edge conduction contribute to the measured resistivity. In the Corbino geometry, no edges connect the source and drain contacts, eliminating edge contributions from conduction measured with transport. While insensitive to the Hall voltage, transport in Corbino devices is an intensive measurement of the bulk conductivity $\sigma_{xx}$, which is invaluable when studying dissipation in systems exhibiting topological edge transport \cite{Hata2016,Yokoi1998,Eber1983,Chida2014}.

Here we present Corbino geometry measurements of dissipation in the QAH state in a thin-film sample of Cr-BST. By comparing transport in Corbino and Hall geometries we conclude that dissipation in the QAH state is dominated by bulk processes at all values of temperature and electric field.

\section{Results and Discussion}
\begin{figure}[t!]
\includegraphics[width=\columnwidth]{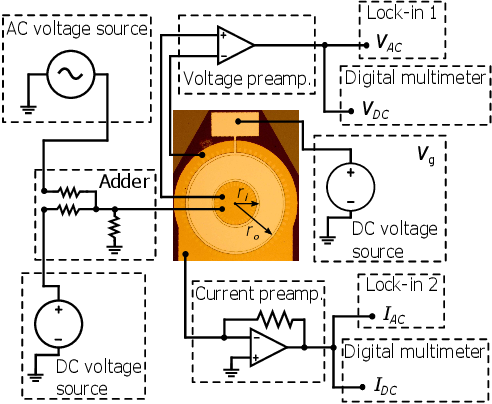}
\caption{\label{fig:circuit}  Measurement schematic and micrograph of device $C_1$ ($r_i$ = 100 $\upmu$m, $r_o$ = 200 $\upmu$m. AC+DC voltage addition in combination with a quasi-four-terminal probe configuration eliminates in-line resistance and allows simultaneous measurement of source-drain voltage and current ($V_{DC}$ and $I_{DC}$) as well as their differential components ($V_{AC}$ and $I_{AC}$). A potential $V_g$ was used to modulate the top gate.}
\end{figure}

A drawback of the Corbino geometry is non-uniform distribution of radial electric fields across the width of the annulus. Here the Corbino width is defined as $W_{\textrm{Corbino}} = r_o - r_i$, the difference between the outer contact radius  $r_o$, and inner contact radius $r_i$. The electric field becomes less uniform as the disk becomes wide compared to the radius of the inner contact (i.e. outside the limit $W_{\textrm{Corbino}} \ll r_i$). To gain insight into how this non-uniformity affects transport, the three devices, $C_1$, $C_2$ and $C_3$, were chosen to have a fixed width of 100 $\upmu$m with varying inner and outer radii ($C_1$: $r_i$ = 100 $\upmu$m, $r_o$  = 200 $\upmu$m. $C_{2,3}$: $r_i$ = 200 $\upmu$m, $r_o$  = 300 $\upmu$m).  Using the AC+DC voltage biased measurement setup shown in Fig.~\ref{fig:circuit} the longitudinal conductivity, \begin{equation}{\label{eq:sigma}}
    \sigma_{xx} = \frac{\ln\left({r_o/r_i}\right)}{2\pi}\frac{I_{DC}}{V_{DC}}
\end{equation} 
was measured simultaneously with the differential conductivity. Note that we define $\sigma_{xx} =\sigma_{xx}\left(V_{DC}\right)$ as a function of bias, which is not necessarily equivalent to the local conductivity, specifically when the $I$-$V$ characteristic is non-linear.

\noindent Three Corbino devices were fabricated from a 6-QL (Cr$_{0.12}$Bi$_{0.26}$Sb$_{0.62}$)$_2$Te$_3$ film grown by molecular beam epitaxy. A chip from the same wafer was used in 2018 to fabricate Hall bars which demonstrated part-per-million quantization of the Hall resistance \cite{Fox2018}. In both cases nearly identical process flows were used (full fabrication details can be found in Section I of the supplemental material). 

\subsection{\label{sec:Init Char}Initial characterization}

\begin{figure*}[t!]
\begin{centering}
\includegraphics[width=5.5in,height=5in,keepaspectratio]{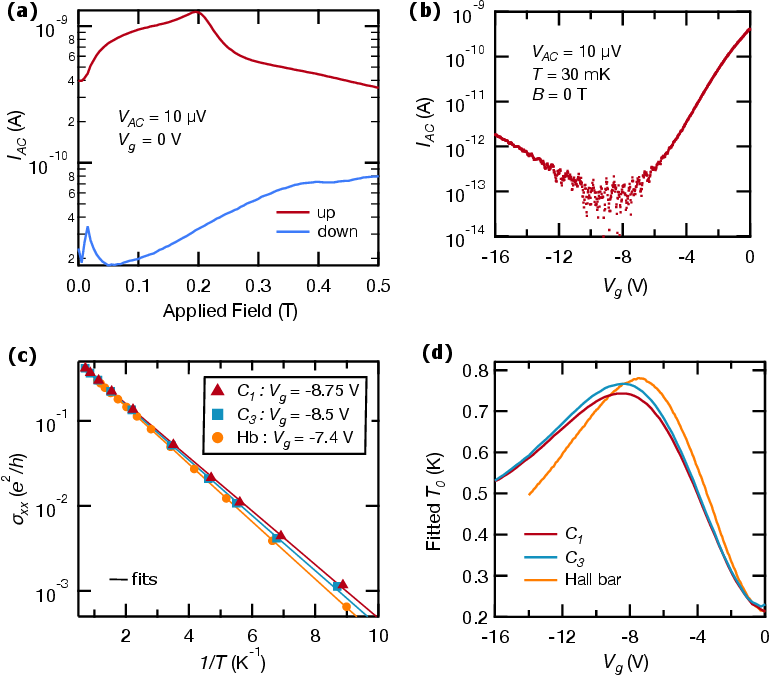}
\caption{\label{fig: 2 init char}{Initial characterization of Corbino disk devices. (a) Lock-in measurement of source-drain current ($I_{AC}$) in device $C_1$ during initial magnetization. $I_{AC}$ increases up to the coercive field of our Cr-BST sample ($B_c \approx 0.2 $ T) and then quickly falls, ultimately dropping below the unmagnetized zero-field value. $I_{AC}$ remains below this initial value while the field is brought back to zero, likely due to orienting of domains so the exchange gap has a single sign everywhere. Prior to returning to zero-field, time was given to allow the sample to return to base temperature after being heated to approximately 60 mK by sweeping the magnet, resulting in the difference in $I_{AC}$ at $B$ = 0.5 T between the upward and downward sweeps. (b) Current measured in device $C_1$ as a function of top gate voltage post-magnetization. In the range ${-14 \;\mathrm{V}} \leq V_g \leq {-4.5 \;\mathrm{V}}$, $I_{AC} < 1$~pA ($R>10$~M$\Omega$), indicating that the QAH state is well-formed and $E_F$ lies within the exchange gap. (c) Arrhenius plots of $\sigma_{xx}$ as a function of $1/T$ for devices $C_1$, $C_3$ and a two-square 100-$\upmu$m-wide Hall bar measured in \cite{Fox2018}. Lines show fits to thermal activation (d) Fitted temperature scale $T_0$ for thermally activated conduction $\sigma_{xx} \propto \exp{-T_0/T}$ as a function of $V_g$ for the same devices as in (c).}}
\end{centering}
\end{figure*}
We first verified the presence of the QAH state. In a Hall bar geometry, this is done by measuring a quantized Hall resistance, $\rho_{yx}= h/e^2$  and vanishing longitudinal resistance. The Corbino geometry, however, does not allow for simple Hall measurements~\cite{Mumford2020}. Measurement of a minimum in the radial current, signifying a fully insulating bulk, was instead used to verify QAH. The radial current decreased in two stages - first when the sample was magnetized, opening an exchange gap with spatially uniform sign, and second when the Fermi level was tuned near the center of the exchange gap using the gate.

The sample was cooled to a base temperature of 30 mK in a dilution refrigerator and the source-source drain current, $I_{AC}$, was measured under a 10 $\upmu$V RMS AC bias and zero DC bias. At gate voltage $V_g  = 0$, $I_{AC} \approx 4$ nA initially after cooling the sample from room temperature. Next, the sample was magnetized by applying a perpendicular field $\mu_0$H = 0.5 T and then reducing the field to zero (Fig.~\ref{fig: 2 init char}a). After magnetization with $V_g = 0$,  $I_{AC}  \approx$ 20~pA. Upon tuning $V_g$ (Fig.~\ref{fig: 2 init char}b), we observed further reduction in $I_{AC}$ and a wide range of gate voltage (${-14 \;\mathrm{V}}  \leq V_g \leq {-4.5 \;\mathrm{V}}$), over which $I_{AC}  < $ 1~pA, indicating that the QAH state was well-formed~\cite{Hata2016}. Further investigation demonstrated that heating of the sample and excursions in gate voltage away from this plateau led to partial demagnetization (Section IV of supplement). For this reason, all subsequent measurements were taken under a 500~mT external magnetic field applied perpendicularly to the plane of the sample, to maintain the magnetization.

The Hall bar devices were fabricated in August 2017 while the Corbino devices were fabricated in June 2018. Aging studies have shown that long term exposure of thin-film TIs to atmospheric conditions can destroy the insulating bulk character~\cite{Aguilar2013}. To limit the effects of aging our samples were coated with PMMA after growth, and stored in an N$_2$ dry box prior to device fabrication. After fabrication additional protection was provided by the 40~nm alumina top-gate dielectric.

By comparing conductivity as a function of temperature and gate voltage on the two sets of devices, we confirmed that aging had not led to significant sample degradation. Fig.~\ref{fig: 2 init char}c shows $\sigma_{xx}$ as a function of $1/T$ for devices $C_1$ and $C_3$ as well as the corresponding data for a 100-$\upmu$m-wide Hall bar measured in 2018. For temperatures ${100 \;\mathrm{mK}} < T < {1.2 \;\mathrm{K}}$ , and with $V_g$ tuned within the regime minimizing $I_{AC}$ (${-14 \;\mathrm{V}} \leq V_g \leq {-4.5 \;\mathrm{V}}$), $\sigma_{xx}$ follows thermally activated behavior $\sigma_{xx} \propto e^{-T_0/T}$. Here, conductance was ohmic under the 10~uV RMS AC excitation.\par
The fitted thermal activation temperature scale $T_0$ reaches a maximum of 727~mK at $V^{opt}_g = $ -8.8~V for device $C_1$ and 729 mK at $V^{opt}_g = $  -8.5~V for device $C_3$. While these values are far below what has been seen in scanning tunneling microscopy measurements of a cleaved Cr-BST bulk crystal~\cite{Lee2015} where average exchange gap sizes are ~30 meV, they are remarkably similar to what was measured in the 2018 Hall bar device both in peak value and peak location (Fig.~\ref{fig: 2 init char}d). The similar optimum gate voltages indicates that, despite   fabrication at different times, aging did not significantly shift the native Fermi level between devices; the similarity of the Arrhenius activation scales indicates that the magnetically induced exchange gap did not shrink.

\subsection{\label{sec:SD dep}Source-drain voltage dependence}
\begin{figure}[h t!]
\includegraphics[width=\columnwidth]{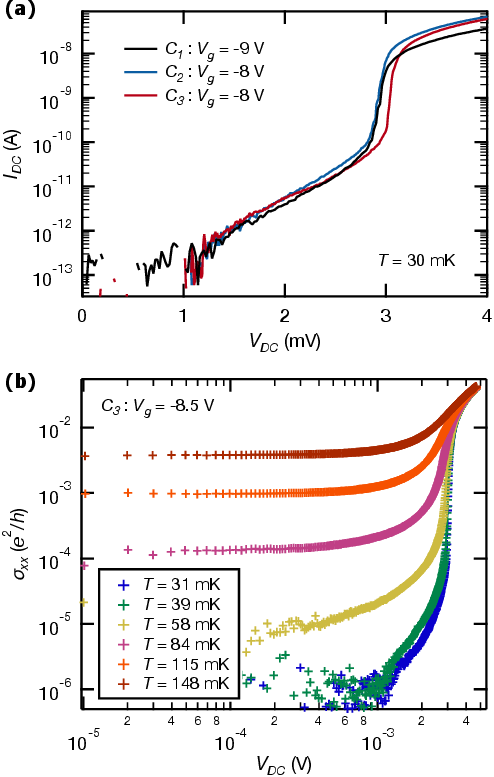}
\caption{\label{fig: 3 SD}{\noindent Source-drain voltage dependence of Corbino devices (all data were taken under an applied external field $B = $ 0.5 T). (a) Measured current $I_{DC}$ as function of applied DC bias voltage $V_{DC}$ for the three Corbino devices, with the gate of each tuned near $V_g = V_g^{opt}$. Raw $I$-$V$ curves for all devices appear similar despite varying geometries. (b) $\sigma_{xx}(V_{DC})$ of device $C_3$ calculated using Eq.~(\ref{eq:sigma}) for various lattice temperatures (as measured by the mixing chamber plate thermometer) while $V_g = V_g^{opt}$. The observed temperature dependence and lack of an ohmic regime at the lowest temperatures is highly consistent with our previous measurements on Hall bar devices~\cite{Fox2018}.}}
\end{figure}

We next considered the behavior of the devices at $V_g = V_g^{opt}$ as a function of source-drain voltage bias, $V_{DC}$. At base temperature no ohmic regime could be observed; we observed highly nonlinear $I$-$V$ characteristics at the lowest currents we could measure ($\sim0.3$~pA). An abrupt increase in dissipation followed at higher bias as the sample underwent breakdown of the QAH state. This behavior was present in all three Corbino devices (Fig.~\ref{fig: 3 SD}a).

Fig.~\ref{fig: 3 SD}b shows $\sigma_{xx}(V_{DC})$ of device $C_3$ at various temperatures. At low temperatures ($T < 50$~mK) a clear transition at breakdown was present. For $T > 50$~mK, ohmic conduction was observed at low bias, and the transition at breakdown was increasingly smeared until a nearly continuous evolution emerged above $\sim 100$~mK. This behavior was consistent with the temperature dependence of breakdown observed in the Hall bar samples (Fig.~3a in~\cite{Fox2018}).

There is no universal method for quantifying the voltage onset of breakdown; even in QH systems, where there is often a discontinuity in the current-voltage characteristic, various criteria have been used to define where breakdown begins~\cite{Eber1983,Nachtwei1999,Yokoi1998,Komiyama1985,Bliek1986}. Here we define the critical breakdown voltage $V_c$ as the source-drain voltage at which $\sigma_{xx}= 5e^{-4}$~$e^2/h$, which roughly corresponds to the dissipation measured in each device when the differential conductivity reaches its peak value at breakdown (Fig.~S3 of supplement) so that the current rises rapidly with increasing voltage (Fig.~\ref{fig: 3 SD}a).

Despite their geometrical differences, we find similar values of $V_c$ in all three Corbino devices with $V_c^{C_1} = 2.9$~mV, $V_c^{C_2} = 2.9$ mV and $V_c^{C_3}  = 3.0$~mV. For a material with uniform conductivity in the Corbino geometry, the radial electric field varies as 
\begin{equation}\label{eq:Er}
E_r(r) = \frac{V_{DC}}{r\ln\left(r_o/r_i\right)}
\end{equation}
and reaches a maximum value of $E_{max}=\frac{V_{DC}}{r_i\ln\left(r_o/r_i\right)}$ at $r_i$, the edge of the inner contact. For a given value of $V_{DC}$, $E_{max}$ is 17\% larger in $C_1$ than in $C_2$ and $C_3$. If breakdown occurred when the maximum electric field in the QAH insulator reached a critical value, we should therefore observe $V_c^{C_{2,3}}$ larger than $V_c^{C_1}$ by 17\%. This was not the case: $V_c^{C_{2,3}}$ were roughly equal to $V_c^{C_1}$, and the $I$-$V$ characteristics of the three devices were similar throughout the breakdown transition without normalizing for device geometry (we emphasize that all devices had equal $W_{\textrm{Corbino}}$).
 
The similarity in unnormalized $I$-$V$ characteristics between the devices suggests that even if differences in radial field distributions are playing some role in the fine details of $\sigma_{xx}$ at breakdown, it appears to be the average electric field 
\begin{equation}\label{eq:Eavg}
    E_{avg} = \frac{V_{DC}}{W_{\textrm{Corbino}}}
\end{equation}
across the annulus that is most relevant for breakdown, at least for the two Corbino geometries used here. The extent to which this finding holds in general for QAH Corbino disks will be the topic of future work; the limited literature on breakdown in QH Corbino geometries does not show this~\cite{Hata2016,Yokoi1998,Bykov2012}. 

\subsection{\label{sec:direct comp}Direct comparison to Hall bar data}

\begin{figure*}[t!]
\includegraphics[width=\textwidth]{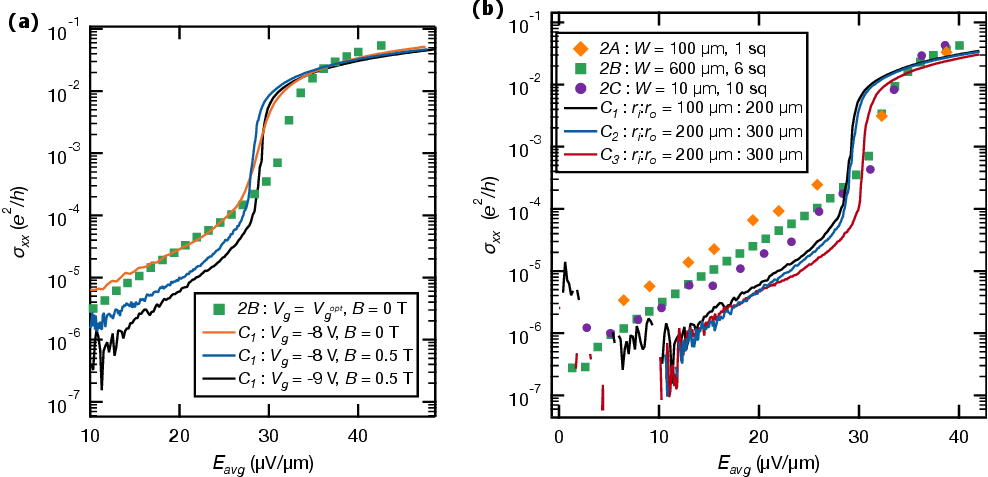}
\caption{\label{fig: 4 comp}{Direct comparison of Corbino and Hall bar transport data. (a) $\sigma_{xx}$ as a function of $E_{avg}$ for Hall bar device $2B$ and Corbino device $C_1$. Hall bar device data were taken at zero field with the gate tuned to optimum. The zero field low bias behavior of device $C_1$ is nearly identical to that of device $2B$. Application of the 0.5 T field reduces low-bias conductivity by roughly a factor of 3 but does not strongly affect the dependence of conductivity on bias. Modest further reduction of $\sigma_{xx}$ is seen as the gate of the device is tuned near optimum $V_g^{opt} = -8.75$ V for device $C_1$. (b) $\sigma_{xx}$ as a function of $E_{avg}$ for three Hall bar devices of varying size (measured in \cite{Fox2018}) and all three Corbino devices. The gate of each device was tuned near the respective optimum.  All devices were fabricated from the same wafer of material. The extreme similarity of the data among the Hall bars and separately among the Corbinos over the full range of bias indicates that bulk conduction dominates dissipation before, during, and after breakdown of the QAH state. The apparent difference between the two sets of devices may be explained by the magnitudes of the magnetic field when the data were taken: zero for the Hall bars versus 0.5 T for the Corbino devices [see panel (a)].}}
\end{figure*}

We want to compare transport in the Hall and Corbino geometries as a function of a single bias parameter. In the QAH regime, the Hall angle $\theta_H \approx 90^{\circ}$ and the electric field is orthogonal to the direction of dissipationless current flow~\cite{Eber1983}. In a Hall bar this electric field corresponds to the transverse field $E_y$, and in a Corbino disk, to the radial (source-drain) field $E_r$. Because we cannot directly probe the local value of electric field, we must appropriately scale extensive measurements to correct for differences in geometry and measurement configuration (current versus voltage bias). As was done in \cite{Fox2018}, we assume for this purpose a uniform distribution of current in the bulk for the Hall bar devices.

Under this assumption, we calculate the average transverse electric field $E_{avg}$ in the Hall geometry as $E_{avg} = E_y = \rho_{yx} j_x  \approx h/e^2 j_x$. Here $j_x$ is the average current density $j_x = I_x/W_{hb}$ in a Hall bar of width $W_{hb}$ biased with current $I_x$. Note that the approximations $\theta_H\approx 90^{\circ}$ and $\rho_{yx}  \approx h/e^2$ hold even after breakdown: for the largest value of current bias used in this comparison $\rho_{yx}$ remained quantized to within 1\% of $h/e^2$, and $\theta_H \geq 80^{\circ}$~\cite{Fox2018}. In the Corbino geometry, in place of $E_r$, which varies as a function of radius, we use $E_{avg}$ of Eq.~(\ref{eq:Eavg}). Strictly speaking $E_{avg}\approx E_r$ only in the limit of a narrow disk ($W_{\textrm{Corbino}} \ll r_i$), but the similar $I$-$V$ characteristics of $C_1$ and $C_{2,3}$ (Fig.~\ref{fig: 3 SD}a) justify using $E_{avg}$ as a representative parameter. Fine details regarding the dependence of $\sigma_{xx}$ on local electric field will require a systematic study of Corbino disks over a broader range of dimensions, and will be the topic of future work.

Figs.~\ref{fig: 4 comp}a and \ref{fig: 4 comp}b show transport data for Corbino devices $C_1$, $C_2$, and $C_3$ and Hall bar devices $2A$, $2B$, and $2C$ (measured in \cite{Fox2018}) as a function of $E_{avg}$. For the Corbino devices, $\sigma_{xx}$ was calculated using Eq.~(\ref{eq:sigma}). For the Hall bar devices $\sigma_{xx} = \rho_{xx}/(\rho_{xx}^2 + \rho_{yx}^2)$; $\rho_{yx}$  was measured in DC using a commercial cryogenic current comparator system, and $\rho_{xx}$ directly using the system's nanovoltmeter as described in \cite{Fox2018}. The input impedance of the nanovoltmeter is $\sim 100$~G$\Upomega$, so leakage currents were insignificant in these measurements \cite{Delahaye2003}.

As previously mentioned, due to partial demagnetization all Corbino disk data presented thus far were taken under a 0.5~T perpendicular magnetic field. This was not the case for the Hall bar data presented in \cite{Fox2018}], which were taken with zero external magnetic field. As seen in Fig.~\ref{fig: 4 comp}a, application of the magnetic field reduced $\sigma_{xx}$ by roughly a factor of 3 in the prebreakdown regime. Modest further reduction was observed as $V_g$ was tuned to $V_g=V_g^{opt}$. Whether demagnetization has a significant effect on transport in Hall bar devices will be the topic of future work~\cite{Lachman2015}. The analysis that follows will focus on the overall shape of $\sigma_{xx}(E_{avg})$, which appears unaffected by the presence of the external magnetic field (Section IV of supplement).

\subsection{\label{sec:prebreakdown}The prebreakdown regime}
If dissipation in the prebreakdown regime were due solely to backscattering in quasihelical edge states, no current would flow between source and drain in the Corbino geometry. Our measurements show that this is not the case. Not only does current flow through the Corbino devices within the prebreakdown regime, but the dependence of $\sigma_{xx}$ on $E_{avg}$ in the three Corbino devices is nearly identical to that in the three Hall bars (Fig.~\ref{fig: 4 comp}b). In all devices we observe an exponential dependence of $\sigma_{xx}$ on $E_{avg}$. Fitting these low bias data to the function $\sigma_{xx} = \sigma_0\exp{\left(aeE_{avg}/k_BT\right)}$, where $e$ is electron charge and $T$ is our 30 mK base temperature, yields $a = (600 \pm 60)$~nm for all six devices (fitting parameters are presented in Table~I of the supplemental material). The parameter $a$ reflects a length scale within the field-assisted thermal activation~\cite{Shimada1998} and non-ohmic variable range hopping~\cite{Pollak1976} models, both of which are consistent with the observed exponential behavior and are discussed extensively in our previous work \cite{Fox2018}. 

Since the conductivity scales with electric field in the same way in Hall and Corbino device geometries, we infer that its dominant contributing process is shared between both. In the Hall geometry, both edge and bulk processes could reasonably be linked to the measured dissipation. But in the Corbino geometry, edge processes do not contribute at all. Therefore, we argue that dissipative conduction observed in thin film Cr-BST at low temperatures is bulk dominated. If quasihelical modes exist, they are strongly localized, or else their conductance is insignificant compared to that of other dissipative channels. Our findings are inconsistent with the conclusions of a previous study of a similar Cr-BST film~\cite{Cuizu2015}.

The exchange gap in QAH materials is expected to be of order tens of meV, but the observed transport gap is closer to 1~K~\cite{Bestwick2015,Fox2018,Chang2015,Gotz2018,Grauer2015,Feng2015}. Our Corbino geometry measurements show that quasihelical edge modes cannot explain this disparity. The disparity may instead result from charge fluctuations spatially modifying the Fermi level~\cite{Chang2016}, regions with a locally reduced exchange gap due to magnetic disorder, or incomplete localization of midgap defect states.
\subsection{\label{sec:above breakdown}At and above breakdown}
For $E_{avg} \gtrapprox 30$~$\upmu$V/$\upmu$m, $\sigma_{xx}$ begins to rise even more rapidly than the exponential dependence described above, and breakdown occurs. Based on the linear scaling of critical current $I_c$ with Hall bar width, it was concluded in 2018 by some of the present authors that breakdown of the QAH phase in this material likely occurred through bulk conduction~\cite{Fox2018}. The Corbino disk measurements presented here reinforce this conclusion, removing possible ambiguity.

In both geometries, $\sigma_{xx}$ sharply increases by nearly two orders of magnitude as the QAH state breaks down, and the transition to the highly dissipative state becomes increasingly smeared out as the samples are heated above base temperature (Fig.~\ref{fig: 3 SD}b and Fig.~3b in \cite{Fox2018}). 
Section \ref{sec:SD dep} introduced a phenomenological definition of breakdown based on $I$-$V$ curves, and noted that the value of $E_{avg}$ at breakdown in the Corbino devices appears consistent across multiple devices. We find that this critical field $E_c =$~29--30~$\upmu$V/$\upmu$m. In the Hall bars $E_c = 31$~$\upmu$V/$\upmu$m is strikingly similar. The qualitative and quantitative consistency of breakdown phenomenology between the two device geometries  confirms that breakdown occurs through bulk conduction. This finding echoes studies of QH breakdown incorporating Corbino devices, in which breakdown has been found to be an intrinsic property of the electron system~\cite{Eber1983,Yokoi1998,Komiyama2000,Hata2016}.

Far beyond breakdown, ($E_{avg} > 35$ $\upmu$V/$\upmu$m), the conductivity measured in the Corbino devices falls somewhat below that in the Hall bar devices (Fig.~\ref{fig: 4 comp}a shows the most direct comparison under zero field, while Fig.~\ref{fig: 4 comp}b includes a larger set of devices). A possible contributor to this difference is that Corbino measurements, being two-terminal, are sensitive to contact resistance, which reduces the conductivity inferred from measurements~\cite{Nomokonov2019}.

\section{Conclusions}
We have shown that dissipation in Cr-BST is dominated by bulk processes both before and after breakdown of the QAH state. Furthermore, both dissipation in the prebreakdown regime and the trigger for breakdown are largely a function of average transverse electric field in the devices studied here.

By comparing the conductivity in Hall and Corbino geometries, we have ruled out nonchiral edge modes as a substantial source of dissipation in the Cr-BST thin films we measured, though we cannot rule out their existence altogether. Our comparative measurements protect against important sources of artifacts in sensitive studies of dissipation. First, the Corbino geometry directly probes $\sigma_{xx}$ in the material's two-dimensional bulk. Since our Corbino and Hall measurements show the same dissipation this must occur in the bulk, not on the edges. Second, Corbino measurements do not suffer from systematic offsets in $\sigma_{xx}$ caused by leakage through voltage preamplifiers, which is known to plague current-biased Hall bar measurements~\cite{Delahaye2003,Fischer2005}. Third, given available current-to-voltage preamplifiers (for Corbino measurements) and voltage preamplifiers (for Hall bars), the Joule power at the measurement noise floor is about five orders of magnitude lower for the Corbino configuration. The methodology we have established here can be used to determine the nature of dissipation in other QAH materials, including V-BST~\cite{Grauer2015,Chang2015}, Cr-BST/BST heterostructures~\cite{Mogi2015}, and MnBi$_2$Te$_4$~\cite{Kagerer2020,Deng2020}, a crucial step in the development of improved QAH materials. This method can also be applied to glean a precise understanding of the relationship between bulk and edge transport in other systems, including three-dimensional topological insulator films (expected to have sidewall conduction in parallel with 2D bulk top and bottom surface conduction), quantum spin Hall materials, and higher-order topological insulators.

\section*{Supplementary Material}
\noindent See supplementary material for full details regarding device fabrication, instrumentation, choice of breakdown threshold voltage $V_c$, preamplifer leakage currents in the Hall bar geometry, fitting parameters, and partial demagnetization of the Cr-BST thin film.

\section*{Data Availability}
\noindent The data that support the findings of this study are available from the corresponding author upon reasonable request.
\section*{Acknowledgements}
\noindent We thank George R. Jones and Randolph E. Elmquist for their help in acquiring the 2018 data referenced throughout this report, and Molly P. Andersen and Steven Tran for helpful discussions. Research supported by the U.S. Department of Energy (DOE), Office of Science, Basic Energy Sciences (BES), under Contract DE-AC02-76SF00515 (device fabrication, measurements, and analysis). P.Z., L.P., and K.L.W. acknowledge support from the Army Research Office under Grants No. W911NF-16-1-0472 and No. W911NF-15-1-0561:P00001, and from the National Science Foundation under ERC-TANMS for material synthesis and material characterization. Measurement infrastructure was funded in part by the Gordon and Betty Moore Foundation through grant GBMF3429 and grant GBMF9460, part of the EPiQS Initiative. During the early stages of this work, L.K.R. was supported by the National Science Foundation Graduate Research Fellowship under Grant No. DGE-1656518. I.T.R. acknowledges support from the ARCS Foundation. Part of this work was performed at the Stanford Nano Shared Facilities (SNSF), supported by the National Science Foundation (NSF) under Award No. ECCS-1542152.

\setlength\bibitemsep{0pt}
\printbibliography

\onecolumn
\begin{center}
\large\textbf{Supplementary Material}
\end{center}
\setcounter{figure}{0}
\setcounter{section}{0}
\renewcommand{\thefigure}{S\arabic{figure}}
\renewcommand{\theequation}{S\arabic{equation}}

\section{Methods}
\subsection{Film growth and device fabrication}\label{sec: sup growth and fab}
The quantum anomalous Hall (QAH) insulator sample studied in this was work was a high quality, single-crystalline 6-quintuple-layer Cr-doped (Bi$_x$Sb$_{1-x}$)$_2$Te$_3$ thin film grown by molecular beam epitaxy (MBE). The growth method was the same as those outlined in \cite{Kou2014} and \cite{Bestwick2015}. A full summary of these growth methods can be found in the supplemental material of our previous work \cite{Fox2018} which used the same wafer of material to fabricate Hall bar devices.\par
As well as being fabricated form the same wafer of material, the Corbino disk devices used in this work were fabricated with nearly identical processes flows as those used to fabricate the Hall bars in \cite{Fox2018}. The only difference in fabrication procedure was in the final step where a patterned selective etch was used to remove excess gate dielectric rather than a full area etch. This was done to reduce the risk of gate failure which is a common problem in Corbino devices due to the fact that a small area of the gate must run across the length of the outer contact as shown by the highlighted region in Fig. S1. The fabrication description that follows is a summary of the detailed procedure outlined in the supplemental material of our previous work \cite{Fox2018}.

\begin{figure}[H]
\begin{centering}
\includegraphics[keepaspectratio,width=2.5in,height=2.5in]{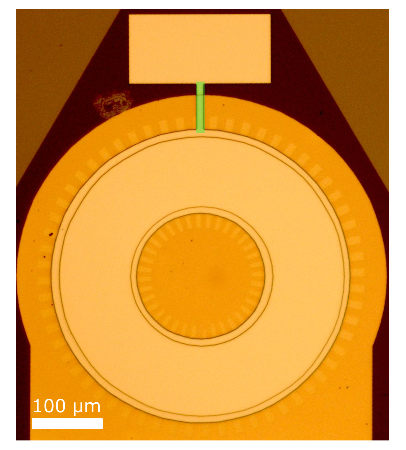}
\caption{\label{fig: sup micrograph} Micrograph of Corbino device $C_1$ after fabrication but prior to wire bonding. The false colored area is used to emphasize the fragile region of the top gate. }
\end{centering}
\end{figure}

Using contact photolithography and Ar ion milling, the mesa of Corbino device $C_1$ ($C_2$ and $C_3$) was defined with $r_i$ = 100 $\upmu$m and $r_o$ = 200 $\upmu$m ($r_i$ = 200 $\upmu$m and $r_o $= 300 $\upmu$m). 5 nm / 100 nm Ti / Au ohmic contacts were deposited by e-beam evaporation before growing a 40-nm Al$_2$O$_3$ gate dielectric over the entire sample using atomic layer deposition. Photolithography and e-beam evaporation were again used to define and deposit 5 nm / 80 nm Ti / Au gate electrodes. Finally, a selective etch was performed to remove the dielectric covering the contact pads. Fig. \ref{fig: sup micrograph} shows an optical micrograph of device device $C_1$ after the full process described above was completed but prior to wire bonding.

\subsection{Instrumentation and Measurement Details}\label{sec: instrumentation}
Transport measurements were performed using the AC+DC voltage biased measurement scheme presented in Fig. 1 of the main text. The bias signals consisted of a DC voltage, $V_{DC}$, and AC differential voltage, $V_{AC}$. The AC and DC bias signals were summed and scaled using voltage dividers. The AC signal was scaled by a factor of 1E-4 and the DC signal by a factor of 1. The summed bias signal was applied at the inner contact of the Corbino device while the outer contact was connected to the virtual ground set by an Ithaco 1211 current preamplifier (chosen to have a gain of $10^7$ V/A, at which its input impedance is 200~$\Omega$) that was used to measure the current across the device. The current The 10 $\upmu$V RMS AC voltage signal was supplied at 13.773 Hz by a Stanford Research Systems SR830 lock-in amplifier. Higher frequencies resulted in significant phase differences between the measured signals and the excitation. The DC voltage was supplied by a Yokogawa 7651 DC source. The potential difference between the inner and outer contact was measured with an NF Corporation LI-75A voltage preamplifier, which has a gain of 100 and a 100 M$\Omega$ input impedance. The DC output of the preamplifier was measured using an Agilent 34401A digital multimeter, and the AC output was measured using a Stanford Research Systems SR830 lock-in amplifier. The DC output of the current preamplifier, $I_{DC}$, was measured using an Agilent 34401A digital multimeter, and the differential current $I_{AC}$ was measured using a Stanford Research Systems SR830 lock-in amplifier. A voltage was applied to the gate electrode using a Yokogawa 7651 DC source.

\subsection{Leakage current through preamplifiers}

It is well known in QH literature that measurements of the longitudinal resistance $R_{xx}$ are impacted by the finite input impedance of the voltmeter used to make the measurement~\cite{Delahaye2003}. Specifically it has been seen that, at $\nu =\pm 1$, $R_{xx}$ gains an additional component, $Z^{res} \approx \frac{R_{xy}^2}{R_{in}} = \frac{h}{e^2}\frac{1}{R_{in}}$ (where $R_{in}$ in the voltmeter input impedance), for one chirality but not the other. This can lead to false assumptions of an additional parallel conduction channel~\cite{Fischer2005}. The asymmetry results from the fact that the finite input impedance provides an additional path to ground and, while quantized, the low-potential side of the device is necessarily at the same potential as the nearly-grounded drain lead. On the other hand, the high-potential side of the device is in equilibrium with the source lead and therefore has a larger potential difference with respect to ground. Below we use the Landauer-B\"uttiker formalism to derive how the longitudinal resistance measured for either chirality is impacted by this leakage current.

\begin{figure}
\begin{centering}
\includegraphics{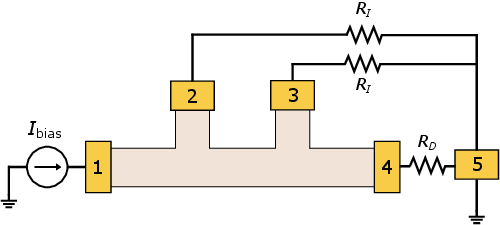}
\caption{\label{fig: sup LB} Example schematic for a current-biased four-terminal measurement. The bias current $I_{\textrm{bias}}$ is injected into the device via terminal 1. The current drains from contact 4 to true ground (contact 5). The total impedance to ground, which includes current preamplifier impedance and line resistance is given by $R_D$. Measurement of $V_{23} = V_{xx}$ is made using a voltage preamplifier  which is assumed to have input impedance of $R_I$.}
\end{centering}
\end{figure}

The Landauer-B\"uttiker formalism states that for a probe sitting at potential $V_i$, the current flowing out of that probe and into the device is given by:

\begin{equation}\label{eq: LB I}
    I_i = \frac{e^2}{h}\sum_j\left(T_{ij}V_i - T_{ji}V_j\right)
\end{equation}
where $T_{ij}$ represents the probability of transmission from terminal $i$ to  terminal $j$.

Here we are assuming that a current $I_{\textrm{bias}}$ is being injected into the device from terminal 1. This current drains out of the device at terminal 4 and into terminal 5 which represents true ground. The input impedance of the voltage preamplifier is given by $R_I$ (inverse $G_I$). $R_D$ (inverse $G_D$) represents the total impedance to ground of the drain terminal (current preamplifier input impedance plus line resistance). 

For magnetization $M = +1$ (counterclockwise chirality) the non-zero off-diagonal transmission coefficients are given by
\begin{equation*}
\begin{split}
    &T_{12} = T_{23} = T_{34} = T_{41} = 1\\
    &\frac{e^2}{h}T_{25} = \frac{e^2}{h}T_{52} = \frac{e^2}{h}T_{35} = \frac{e^2}{h}T_{53} = G_I\\
    &\frac{e^2}{h}T_{45} = \frac{e^2}{h}T_{54} = G_D
\end{split}
\end{equation*}

Using Eq. (\ref{eq: LB I}) and letting $G_0 = \frac{e^2}{h}$ we can write:
\begin{equation} \label{eq: LB matrix}
    \begin{pmatrix}
    I_1\\
    0\\
    0\\
    0\\
    I_5
    \end{pmatrix}
    = 
    \begin{bmatrix}
    G_0 & 0 & 0 & -G_0& 0\\
    -G_0 & G_0+ G_I & 0 & 0 & -G_I\\
    0 & -G_0 & G_0+G_I & 0 & -G_I\\
    0 & 0 & -G_0 & G_0+G_D& -G_D\\
    0 & -G_I & -G_I & -G_D & 2G_I+G_D
    \end{bmatrix}
     \begin{pmatrix}
    V_1\\
    V_2\\
    V_3\\
    V_4\\  
    V_5
    \end{pmatrix}  
\end{equation}

\noindent Keeping in mind that $I_1 = I_{\textrm{bias}}$ and that Kirchhoff's junction rule still applies, we know that $I_5 = -I_1 = -I_{\textrm{bias}}$. Furthermore, we can choose true ground as our reference potential and set $V_5 = 0$ to write

\begin{equation} \label{eq: LB matrixinverse}
    \begin{pmatrix}
    V_1\\
    V_2\\
    V_3\\
    V_4\\
    \end{pmatrix}
    = 
    \begin{bmatrix}
    G_0 & 0 & 0 & -G_0\\
    -G_0 & G_0+ G_I & 0 & 0 \\
    0 & -G_0 & G_0+G_I & 0 \\
    0 & 0 & -G_0 & G_0+G_D\\
    \end{bmatrix}^{-1}
     \begin{pmatrix}
   I_{\textrm{bias}}\\
   0\\
   0\\
    0
    \end{pmatrix}  
\end{equation}

\noindent Letting $A = \frac{1}{G_I^2(G_D+G_0)+2G_IG(G_D+G_0)+G_DG_0^2}$ and noting that we only require the first column of the inverted matrix from Eq. (\ref{eq: LB matrixinverse}), we find
\begin{equation} \label{eq: LB matrixinverse2}
    \begin{pmatrix}
    V_1\\
    V_2\\
    V_3\\
    V_4\\
    \end{pmatrix}
    = A
    \begin{bmatrix}
    \frac{G_I^2G_D+G_I^2G_0+2G_IG_DG_0+2G_IG_0^2+G_DG_0^2+G_0^3}{G_0} & \dots \\
    G_IG_D+G_IG_0+G_DG_0+G_0^2 & \dots\\
    G_DG_0+G_0^2 & \dots\\
    G_0^2 & \dots
    \end{bmatrix}
     \begin{pmatrix}
   I_{\textrm{bias}}\\
   0\\
   0\\
    0
    \end{pmatrix}  
\end{equation}

We are interested in the effect of preamplifier leakage on the measured longitudinal resistance 
\begin{equation} \label{eq: Rxx general}
R_{xx}^{\textrm{meas}} = \frac{V_2 - V_3}{I_{\textrm{meas}}}
\end{equation}
where $I_{\textrm{meas}}$ is the current measured by the current preamplifier. As some of the total bias current is lost to ground through the voltage preamplifiers, this current will not be equal to $I_{\textrm{bias}}$ and is instead given by
\begin{equation}
    I_{\textrm{meas}} = \frac{V_4}{R_D}
\end{equation}
Solving for $V_2$ and $V_3$ using Eq. (\ref{eq: LB matrixinverse2}) yields:
\begin{equation} \label{eq: v2}
    V_2 = \frac{G_IG_D+G_IG_0+G_DG+G_0^2}{G_I^2(G_D+G_0)+2G_0G_I(G_0+G_D)+G_DG_0^2}I_{\textrm{bias}}
\end{equation}

\noindent and 

\begin{equation} \label{eq: v3}
    V_3 = \frac{G_0G_D+G^2}{G_I^2(G_D+G_0)+2G_0G_I(G_0+G_D)+G_DG_0^2}I_{\textrm{bias}}
\end{equation} 

\noindent Similarly, we can solve for $I_{\textrm{meas}}$ using Eq. (\ref{eq: LB matrixinverse2}) and noting that $G_D = 1/R_D$. This yields
\begin{equation} \label{eq: Imeas}
    I_{\textrm{meas}} = \frac{G_0^2G_D}{G_I^2(G_D+G_0)+2G_0G_I(G_0+G_D)+G_DG_0^2}I_{\textrm{bias}}
\end{equation}

\noindent By substituting Eqs. (\ref{eq: v2}), (\ref{eq: v3}), and (\ref{eq: Imeas}) into Eq. (\ref{eq: Rxx general}), we can solve for the longitudinal resistance ($R_{xx}^{\textrm{meas(+)}}$) measured in the counterclockwise chirality ($M = +1$). Doing so yields
\begin{equation} \label{eq:Rxx plus}
    R_{xx}^{\textrm{meas(+)}} = \frac{G_I(G_0+G_D)}{G_0^2G_D}\\
\end{equation}

A similar procedure can be followed for the clockwise chirality ($M=-1$). The general matrix equation is given by:

\begin{equation} \label{eq: LB matrixinverse3}
    \begin{pmatrix}
    V_1\\
    V_2\\
    V_3\\
    V_4\\
    \end{pmatrix}
    = 
    \begin{bmatrix}
    G_0 & -G_0 & 0 & 0\\
    0 & G_0+ G_I & -G_0 & 0 \\
    0 & 0& G+G_I & -G_0 \\
    -G_0 & 0 & 0 & G_0+G_D\\
    \end{bmatrix}^{-1}
     \begin{pmatrix}
   I_{\textrm{bias}}\\
   0\\
   0\\
    0
    \end{pmatrix}  
\end{equation}

\noindent Solving for the measured value of the longitudinal resistance $R_{xx}^{\textrm{meas(-)}}$ yields

\begin{equation} \label{eq:Rxx minus}
    R_{xx}^{\textrm{meas(-)}} = \frac{G_0^2-(G_IG_0 + G_0^2)}{G_D(G_I^2+2G_IG_0+G_0^2)}
\end{equation}

As an example, we calculate $R_{xx}^{\textrm{meas(+)}}$ and $R_{xx}^{\textrm{meas(-)}}$ using the input impedances of the instrumentation used in this experiment, $R_I = 100$~M$\Upomega$ and $R_D = 1.4$~k$\Upomega$ (which is the series resistance of the current preamplifier's $200~\Upomega$ input impedance and the $1.2$~k$\Upomega$ resistance of the measurement line). Plugging these values into Eqs. (\ref{eq:Rxx plus}) and (\ref{eq:Rxx minus}) yields: $R_{xx}^{\textrm{meas(+)}} \approx 7$~$\Upomega$ and $R_{xx}^{\textrm{meas(-)}} \approx -0.3$~$\Upomega$.

From the above analysis it can be seen that the measured longitudinal resistance can deviate significantly from its near zero value, particularly when $R_{xx}$ is measured on the high-potential side of the device. Similar effects can occur in non-local measurement configurations and thereby complicate analysis or obscure results. On the other hand, the voltage biased Corbino measurement used in this study does not suffer from the same effects. The leakage through the voltage preamplifier occurs at the source terminal while the current is measured at the drain, which is held at near-ground. For this reason, the current flowing across the device is the indeed the measured current. Therefore, measurements of longitudinal conductivity made within the range accessible to our instrumentation accurately reflect $\sigma_{xx}$ of the device. 

\section{Choice of $V_c$}\label{sec: choice of Vc}
\begin{figure}[H]
\begin{centering}
\includegraphics[keepaspectratio,width=3.5in,height=3.5in]{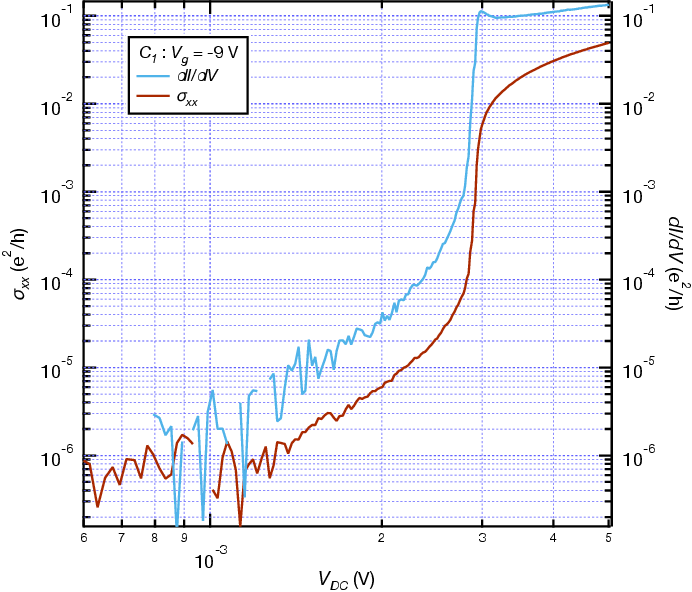}
\caption{\label{fig: sup didv} longitudinal conductivity, $\sigma_{xx}(V_{DC})$,  and differential conductivity, $\frac{dI}{dV}$ measured in device $C_1$ as a function of $V_{DC}$ at base temperature and with the gate of the device tuned near optimum ($T$ = 30 mK and $V_g \approx V_g^{opt}$). $\frac{dI}{dV}$ reaches a local maximum at $V_{DC} = $ 29 mV while $\sigma_{xx} \approx 0.5 \times 10^{-4}$ $e^2/h$}
\end{centering}
\end{figure}

 One of the most common methods used to define the threshold current (or voltage) for breakdown is via the use of a cutoff. Using such a definition, one defines the onset of breakdown as the bias at which the measured dissipation in the system reaches a threshold value. The cutoff value can either be arbitrary~\cite{Bliek1986} or have some physical motivation~\cite{Nachtwei1999,Gotz2018}. The cutoff value  chosen in this report  ($\sigma_{xx} = 0.5 \times 10^{-4}$ $e^2/h$) was chosen as the value of $\sigma_{xx}$ measured in all three devices (at base temperature and with $V_g \approx V_g^{opt}$) when the current between source and drain increased most rapidly as a function of bias voltage at breakdown  (i.e. the cutoff value chosen roughly corresponds to the value of $\sigma_{xx}$ when $\frac{dI}{dV}$ reached a local maximum at breakdown for $T=30$ mK and $V_g \approx V_g^{opt}$) as seen in Fig. \ref{fig: sup didv} for device $C_1$. It is important to note that the differential conductivity is unbounded after breakdown, so the maximum specifically refers to the local maximum observed just after $\sigma_{xx}$ begins to deviate from the exponential dependence observed in the prebreakdown regime. Additionally, the peak observed in $\frac{dI}{dV}$ was present only at the lowest temperatures; however, the defined cuttoff value for $\sigma_{xx}$ can still be used to define the onset of breakdown, even at elevated temperatures.

The AC+DC measurement scheme allowed for simultaneous measurement of the source-drain current $I_{DC}$ and its differential component $I_{AC}$. Similarly to longitudinal conductivity $\sigma_{xx}(V_{DC})$ (calculated using Eq. (1) of the main text), the differential conductivity $\frac{dI}{dV}\left(V_{DC}\right)$ was calculated as
\begin{equation}\label{eq:didv}
    \frac{dI}{dV} = \frac{\ln(r_o/r_i)}{2\pi}\frac{I_{AC}}{V_{AC}}.
\end{equation}
\section{Exponential fits in the prebreakdown regime}\label{sec: exp fits}
In the main text we noted that the exponential dependence 

\begin{equation}\label{eq: expfit}
    \sigma_{xx} = \sigma_0 e^{aeE_{avg}/k_BT}
\end{equation}
observed in the prebreakdown regime at the lowest temperatures could be consistent with a field assisted thermal activation model or a non-ohmic variable range hopping model. Here the parameter $a$ represents a length scale that may reflect the localization length of disorder-localized mid-gap states or the average hopping distance in our system. The applicability of each of these models to our data was discussed extensively in the main text of our previous work \cite{Fox2018} as well as in the supplemental material of the same work.

Fig. \ref{fig: sup hbcd exp}a shows the longitudinal conductivity of Hall bar devices $2A$-$C$ as a function of $E_{avg}$. This data was taken with the gate of each device tuned near optimum ($V_g \approx V_g^{opt}$) and at zero external magnetic field ($B = 0$ T). In 2018 it was found that fitting of these data to Eq. (\ref{eq: expfit}) for $E_{avg} \leq 26 $ $\upmu$V/$\upmu$m yielded similar values for the parameter $a$ for all three devices (rows 1-3 of Table \ref{tab:table1}) with $a = 600$ nm $\pm$ 10\% for all. The fitted prefactor $\sigma_0$ was less consistent between devices, differing by nearly a factor of 5 between deceives $2A$ and $2C$. We believe some of this discrepancy can be attributed to material inhomogeneities~\cite{Fox2018} as well as partial demagnetization of the devices. In the Corbino devices this partial demagnetization was seen to increase $\sigma_{xx}$ in the low-bias regime (Fig. 4a of the main text). Because the Hall bar data was taken at zero field, it is conceivable that varying degrees of demagnetization could explain not only the spread in the Hall bar data at low bias but the variation in the fitted prefactor, $\sigma_0$, as well.\\
\tab We find the exponential dependence of $\sigma_{xx} $ on $E_{avg}$ in the prebreakdown regime for three Corbino devices, $C_1$, and $C_2$ and $C_3$, to be identical to that observed in the Hall bar devices (rows 4-6 of Table \ref{tab:table1}). Again $a = 600$ nm $\pm$ 10\% for all. Figs. \ref{fig: sup expfits c1 comp}(b)-(d) show the longitudinal conductivity of each of the three Corbino devices as a function of $E_{avg}$ as well as the fits to Eq. (\ref{eq: expfit}) (fit range shown as a solid line and extrapolated values at higher fields is shown by the dashed line.) This data was taken with the gate of each device tuned near optimum ($V_g \approx V_g^{opt}$) and under a 0.5~T external magnetic field ($B = 0.5$ T). It is important to note that due to noise and measurement accessibility constraints, the lower bound for the fit range of the Corbino data is $E_{avg} = 10$ $\upmu$V/$\upmu$m  ($E_{avg} = 12$ $\upmu$V/$\upmu$m for device $C_2$). This was not necessary for the Hall bar data because longitudinal measurements were performed using the current source and high-input-impedance nanovoltmeter of a cryogenic current comparator (CCC) system\cite{Drung2009}, allowing for more precise measurements. Additionally due to the slightly reduced value of $V_c$ observed in devices $C_1$ and $C_2$ the upper bound of the fit was chosen to be $E_{avg} = 25$ $\upmu$V/$\upmu$m as opposed to $E_{avg} = 26$ $\upmu$V/$\upmu$m for device $C_3$. Notably, the fitted value of $\sigma_0$ for the Corbino devices varies by less than a factor of 2 between devices. Given that these data were taken under an applied field of 0.5~T, this finding is consistent with our belief that partial demagnetization is responsible for some of the variation in $\sigma_0$ observed in the Hall bar data (where the gate voltage of each device was modulated under zero magnetic field).
\begin{table}[H]
\caption{\label{tab:table1} Fitted values of $a$ and $\sigma_0$ as well as their respective errors ($\delta a$ and $\delta \sigma_0$). Fit values obtained by fitting the low bias measurements of $\sigma_{xx}$ as a function of $E_{avg}$ for each device shown in Fig. \ref{fig: sup hbcd exp}. Curve fitting was preformed using a nonlinear least squares fit. Detailed descriptions of the fitting procedure and error analysis for the Hall bar data (devices $2A-C$, rows 1-3) are given in the supplemental material of \cite{Fox2018} To account for measurement uncertainty in the case of the Corbino data (devices $C_{1.2,3}$, rows 4-6), the fit errors are assumed to be 10\% (20\%) of the fitted value $a$ ($\sigma_0$)}
{\renewcommand{\arraystretch}{1.3} 
\begin{tabular*}{\textwidth}{c @{\extracolsep{\fill}} ccccc}
\hline \hline
 Device&$a$ (nm)&$\sigma_0$ ($e^2/h$)&$\delta a$ (nm) &$\delta \sigma_0$ ($e^2/h$)\\
\hline 
Hall bar device $2A$& $6.0\times10^{2}$ & $9.0\times 10^{-7}$ & $4.0\times10^1$ & $2.0\times 10^{-7}$ \\
Hall bar device $2B$& $6.0\times10^{2}$ & $3.6\times 10^{-7}$ & $3.0\times10^1$ & $0.1\times 10^{-7}$ \\
Hall bar device $2C$& $6.2\times10^{2}$ & $1.9\times 10^{-7}$ & $6.0\times10^1$ & $0.7\times 10^{-7}$ \\
Corbino device $C_1$& $6.0\times10^{2}$ & $5.7\times 10^{-8}$ & $6.0\times10^1$  & $1.1\times 10^{-8}$ \\
Corbino device $C_2$& $6.2\times10^{2}$ & $4.1\times 10^{-8}$ & $6.2\times10^1$  & $8.2\times 10^{-9}$ \\
Corbino device $C_3$& $5.9\times10^{2}$ & $3.6\times 10^{-8}$ & $5.9\times10^1$  & $7.2\times 10^{-9}$ \\
\hline \hline
\end{tabular*}}
\end{table}

\begin{figure}[H]
\begin{centering}
\includegraphics{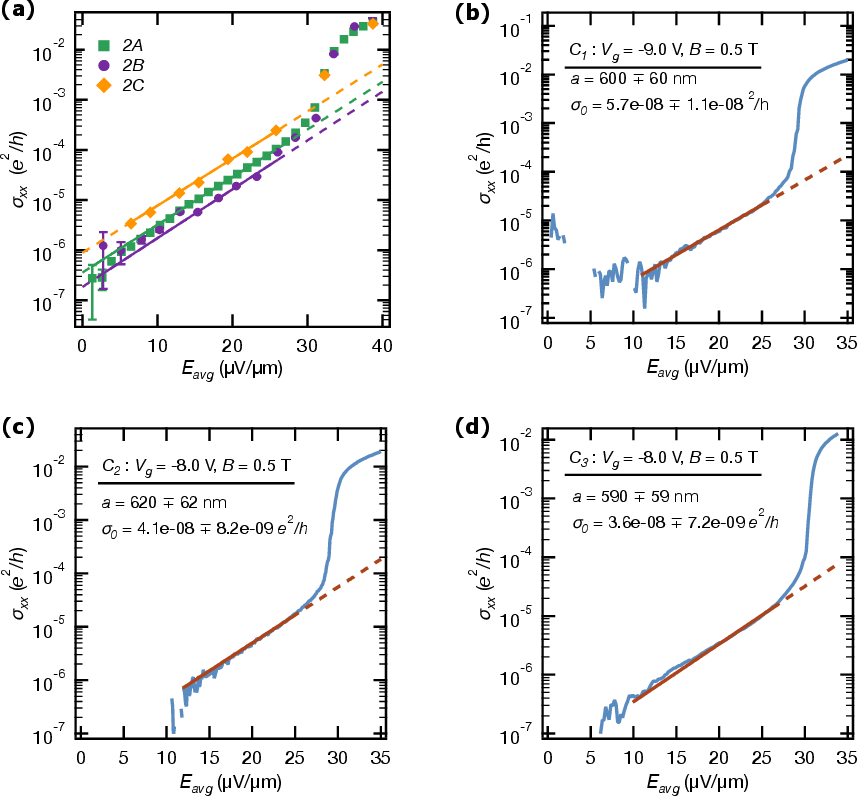}
\caption{\label{fig: sup hbcd exp} Exponential dependence of $\sigma_{xx}$ in the prebreakdown regime. (a) Measurements of $\sigma_{xx}$ as a function of $E_{avg}$ in Hall bar devices $2A$, $2B$ and $2C$ at T = 32 mK. Fits to Eq. (\ref{eq: expfit}) for $E_{avg} < $ 26 $\upmu$V/$\upmu$m (prebreakdown) are show by the solid lines, while extrapolated fits at higher bias are shown with dashed lines. Data adapted from \cite{Fox2018}. (b) Measurements of $\sigma_{xx}$ as a function of $E_{avg}$ in Corbino device $C_1$ at T = 30 mK. Fits to Eq. (\ref{eq: expfit}) for 10 $\upmu$V/$\upmu$m$<E_{avg} < $ 25 $\upmu$V/$\upmu$m (prebreakdown, lower bound chosen to account for noise and measurement accessibility constraints) are show by the solid lines, while extrapolated fits at higher biases are shown with dashed lines. (c) and (d) show similar data for devices $C_2$ and $C_3$ respectively. The fitting ranges for these data were chosen to be 12 $\upmu$V/$\upmu$m$<E_{avg} < $ 25 $\upmu$V/$\upmu$m for device $C_2$ and 10 $\upmu$V/$\upmu$m$<E_{avg} < $ 26 $\upmu$V/$\upmu$m for device $C_3$. Again, these lower bounds were chosen based on measurement constraints and the upper bounds were dictated by the onset of breakdown.}
\end{centering}
\end{figure}

\section{Partial demagnetization}\label{sec: sup demag}
\subsection{Evidence of partial demagnetizing}
\begin{figure}[H]
\begin{centering}
\includegraphics{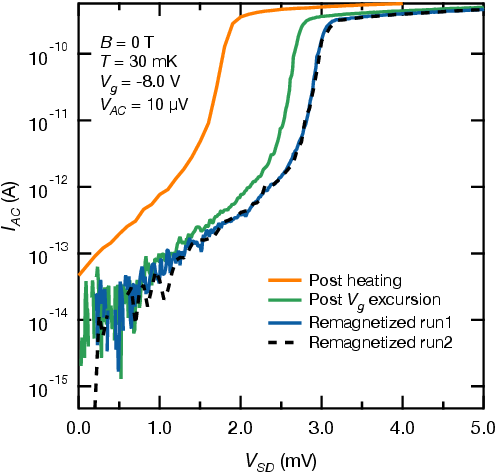}
\caption{\label{fig: sup demag}Partial demagnetization of Cr-BST film. $I_{AC}$ measured as a function of DC bias voltage $V_{DC}$ for device $C_1$ at base temperature ($T = 30 $~mK), $V_g = -8.0$~V and zero external magnetic field ($B=0$~T). Partial demagnetization of the film is indicated by a decrease in breakdown threshold voltage after heating the sample (orange/leftmost curve) or a gate voltage excursion (green/middle curve). Voltage bias sweeps did not cause demagnetization has indicated by the repeatability at zero field (solid blue and dashed black/rightmost curves).}
\end{centering}
\end{figure}

It was noted in the main text that partial demagnetization of the sample was observed upon heating and after gate voltage excursions. This partial demagnetization can be seen in Fig. \ref{fig: sup demag} as a reduction in breakdown threshold bias voltage, referred to as $V_c$ in the main text.

Fig. \ref{fig: sup demag} shows multiple measurements of $I_{AC}$ of device $C_1$ as the DC bias voltage, $V_{DC}$, was swept from 0 V to 5.0 mV. All measurements presented in the figure were taken at base temperature ($T$ = 30 mK) under zero external field ($B$ = 0 T) and with the gate of the device tuned to $V_g$ = -8.0 V. The order and method in which the curves displayed in Fig. \ref{fig: sup demag} were obtained is summarized below:
\begin{adjustwidth}{2em}{2em}
\begin{enumerate}
    \item Sample was cooled to base temperature $T = 30$~mK
    \item Device $C_1$ was magnetized with a $B = 0.5$~T magnetic field
    \item Magnetic field was reduced to $B=0$~T
    \item Sample was heated to $T=1.2$~K and gate voltage was swept from $V_g = -16$~V to $V_g = 0$~V (used to obtain $V_g^{opt}$)
    \item Sample was cooled back to base temperature, $T = 30$~mK
    \item Gate voltage was ramped from $V_g = -16.0$~V  to $V_g = -8.0$~V
    \item Measured $I_{AC}$ vs $V_{DC}$ (leftmost solid [orange] curve)
    \item Sample was remagnetized with a $B = 3.0$~T magnetic field, with the gate voltage fixed at $V_g=-8.0$~V
    \item Magnetic field was reduced to $B =0$~T, with the gate voltage fixed at $V_g=-8.0$~V
    \item Measured $I_{AC}$ vs $V_{DC}$ (rightmost solid curve [blue] curve)
    \item Measured $I_{AC}$ vs $V_{DC}$ (dashed [black] curve)
    \item Gate voltage was ramped to $V_g = -16.0$~V and back to $V_g = -8.0$~V
    \item  Measured $I_{AC}$ vs $V_{DC}$ (middle solid [green] curve)\\
\end{enumerate}
\end{adjustwidth}

This series of measurements suggests that excursions in temperature and gate voltage may cause demagnetization, while sweeping the bias voltage beyond $V_c$ does not. As outlined above, the orange curve (leftmost curve) was taken after the sample was heated to 1.2 K. A large reduction in $V_c$ (compared to $V_c$ of the fully magnetized sample [rightmost, solid blue curve]) was observed, indicating that partial demagnetization had occurred. The solid blue curve (rightmost solid curve) was taken directly after magnetizing the sample while the gate of the device was fixed at $V_g$ = -8.0 V. The same sweep was then immediately repeated (rightmost dashed curve). The repeatability of the data demonstrates that the magnetization of the sample was unaffected by sweeping bias voltage through $V_c$. Finally, The green (middle solid) curve in Fig. \ref{fig: sup demag} shows a similar sweep that was taken after the gate of the device had been swept to -16.0 V and then back to -8.0~V (in zero magnetic field). A moderate reduction in $V_c$ was observed indicating that the gate voltage excursion led to partial demagnetization. Here it is important to note that for all measurements in this work the gate of each device was always tuned by first ramping the gate to -16~V and then ramping to the desired value of $V_g$. By performing gate sweeps over the same range, in the same direction, and at the same speed, we were able to ensure that any observed changes were not a result of hysteresis in the gate. Our observation of gate voltage modulation contributing to partial demagnetization is consistent with previous scanning nanoSQUID (superconducting quantum interference device) magnetic imaging measurements which demonstrated that gate voltage modulation enhances magnetic relaxation in Cr-BST thin film samples\cite{Lachman2015}. \\

\subsection{Effect of external field on transport measurements}\label{sec: sup demag effects}

Fig. 4a of the main text shows that measuring the Corbino devices under the 0.5 T external magnetic field reduced the prebreakdown longitudinal conductivity by roughly a factor of 3. However, the external field does not affect the semilog slope $\sigma_{xx}$ vs $E_{avg}$ (measured by fit parameter $a$) in this regime and therefore does not play a role in our analysis comparing the Hall bar and Corbino transport data. This can be seen by comparing fits of $\sigma_{xx}$ as a function of $E_{avg}$ in the prebreakdown regime for data taken at zero field and under a 0.5 T field while the gate of the device is tuned to the same value. 

\begin{figure}[H]
\begin{centering}
\includegraphics{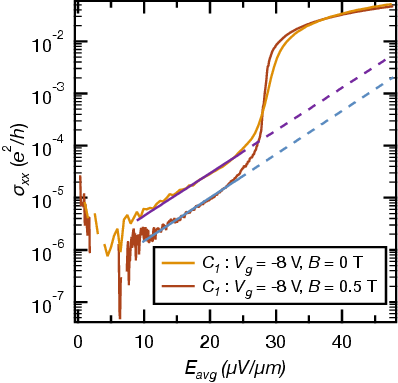}
\caption{\label{fig: sup expfits c1 comp} Measurements of $\sigma_{xx}$ as a function of $E_{avg}$ in Corbino device $C_1$ (T = 30 mK) at $B$ = 0~T and $B$ = 0.5~T.  Fits to Eq. (\ref{eq: expfit}) for 10 $\upmu$V/$\upmu$m$<E_{avg} < $ 26 $\upmu$V/$\upmu$m (prebreakdown regime; lower bound chosen to account for noise and measurement accessibility constraints) are show by the solid lines, while extrapolated fits at higher biases are shown with dashed lines.}
\end{centering}
\end{figure}

Fig. \ref{fig: sup expfits c1 comp} shows  $\sigma_{xx}$ as a function of $E_{avg}$ for device $C_1$ at $B = 0$ T and $B = 0.5$ T both taken with $V_g = -8.0$ V (these traces were taken from Fig. 4a of the main text). Fitting these data to Eq. (\ref{eq: expfit}) over the same range (10 $\upmu$V/$\upmu$m $ < E_{avg} <$ 25 $\upmu$V/$\upmu$m) yields the same vale of $a$ for each as seen in Table \ref{table: C1 exp comp}. 

\begin{table}[H]
\caption{\label{table: C1 exp comp} Fitted values of $a$ and $\sigma_0$ as well as their respective errors ($\delta a$ and $\delta \sigma_0$) for Corbino device $C_1$. Rows 1 and 2 show fit values for the curves shown in Fig. \ref{fig: sup expfits c1 comp} to  to Eq. (S14). The device was measured with the gate tuned to $V_g = -8.0$~V once under and applied magnetic field of $0.5$~T and once at 0~T after the device had been magnetized with a 3.0~T field. Row 3 is provided as a comparison and shows the same data as row 4 of Table \ref{tab:table1} where device $C_1$ was measured under a 0.5~T field with the gate tuned near optimum with $V_g = -9.0$~V. To account for measurement uncertainty, the fit errors are assumed to be 10\% (20\%) of the fitted value $a$ ($\sigma_0$).}
{\renewcommand{\arraystretch}{1.3}
\begin{tabular*}{\textwidth}{c @{\extracolsep{\fill}}ccccccc}
\hline \hline
 Device &$V_g$ (V) &$B$ (T) &$a$ (nm)&$\sigma_0$ ($e^2/h$)&$\delta a$ (nm) &$\delta \sigma_0$ ($e^2/h$)\\
\hline
 $C_1$ & -8.0 & 0 & $4.9\times10^2$ & $6.9\times 10^{-7}$ & $4.9\times10^1$  & $1.4\times 10^{-7}$ \\
 $C_1$ & -8.0 & 0.5 & $4.9\times10^2$ & $2.1\times 10^{-7}$ & $4.9\times10^1$  & $4.2\times 10^{-8}$  \\
 $C_1$& -9.0 & 0.5 & $6.0\times10^{2}$ & $5.7\times 10^{-8}$ & $6.0\times10^1$   & $1.1\times 10^{-8}$ \\
 \hline \hline
\end{tabular*}}
\end{table}

\subsection{Choice of external magnetic field strength}\label{sec: field strength}
Fig. \ref{fig: sup expfits c1 comp} and Table \ref{table: C1 exp comp} demonstrate that the $B= 0.5$~T external magnetic field does affect the scaling of $\sigma_{xx}$ with $E_{avg}$ in the prebreakdown regime. (Specifically, the parameter $a$ for exponential scaling is the same for measurements performed at 0.5~T and 0~T, even though the constant factor $\sigma_0$ varies.) Regardless, it is still important to discuss why an external magnetic field of $B=0.5$~T was chosen rather than a field closer to that of the coercive field $B_c \approx 0.2$~T.

Our goal when measuring the Corbino devices under an applied field  was to ensure that partial demagnetization did not occur when the sample was heated or when the gate voltage was swept. It is true that the QAH effect can be obtained or recovered in our devices, and in other QAH insulators, upon magnetization in an external field comparable to the coercive field. However, previous work on Cr-BST has shown that the magnetization is highly non-uniform at fields near the apparent coercive field; in fact, the magnetization remains non-uniform even at fields of twice the coercivity\cite{Lachman2015}. Magnetization reversal occurs through a series of random events in which isolated nanoscale islands undergo a reversal of their out-of-plane magnetic moment, and the fields at which these events occur may have substantial spread around the average coercive field of the material.

Measuring the Corbino devices in a 0.5 T field, rather than a field nearer the coercive field, prevented partial demagnetization while simultaneously ensuring that the magnetization was as uniform as possible. Furthermore, Fig. \ref{fig: sup expfits c1 comp} demonstrates that while this 0.5~T field was large enough to ensure that partial demagnetization would not occur, it did not affect the slope on a semi-log scale of $\sigma_{xx}\left(E_{avg}\right)$ for $E_{avg} < E_c$.

\section{Linear scale plot of Corbino and Hall bar data }\label{sec: Linear plots}
\begin{figure}[H]
\begin{centering}
\includegraphics{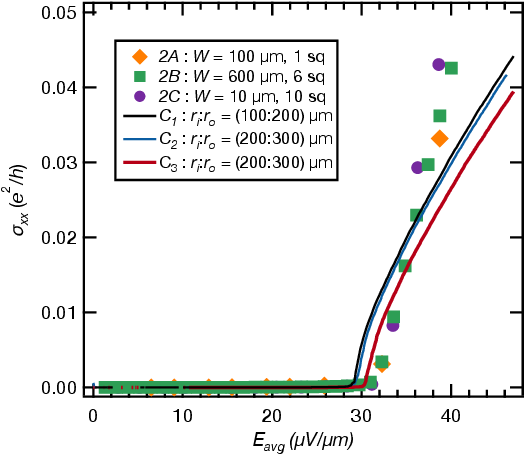}
\caption{\label{fig: sup linear}Linear scale plot of longitudinal conductivity $\sigma_{xx}(V_{DC})$ as a funciton of $E_{avg}$ for the three Corbino geometry devices ($C_1$, $C_2$, and $C_2$) and three Hall bar devices ($2A$, $2B$, and $2C$).}
\end{centering}
\end{figure}

\noindent To further illustrate the difference in $\sigma_{xx}(E_{avg})$ between Corbino and Hall bar geometries in the breakdown regime, Fig.~\ref{fig: sup linear} shows the data of Fig.~4b of the main text plotted on a linear scale. As mentioned in the main text, this difference could be related to the contact resistances picked up in the Corbino measurements, which reduce the inferred conductivity. Because the Corbino geometry is necessarily two-terminal, exact determination of contact resistance is difficult and will be the subject of future work. \\

\end{document}